\documentclass[twocolumn, showpacs,aps,superscriptaddress,prd,notitlepage,showkeys,nofootinbib]{revtex4-1}

\usepackage[normalem]{ulem}
\usepackage{multirow}
\usepackage{amssymb}
\usepackage{amsmath}
\usepackage{graphicx}
\usepackage{dcolumn}
\usepackage{hyperref}
\usepackage{parskip}
\hypersetup{colorlinks,urlcolor=cyan,citecolor=cyan,linkcolor=cyan}
\usepackage{color,units}
\usepackage[dvipsnames]{xcolor} 
\usepackage{lineno}
\usepackage{xspace}
\usepackage{longtable} 
\usepackage{float}  
\usepackage{aas_macros}
\usepackage{amsfonts,wasysym,epsfig,verbatim,subfigure,bm,mathrsfs,lipsum}

\usepackage{color}
\usepackage{hyperref}
\usepackage{amsmath}
\usepackage{multirow}
\usepackage{graphicx}
\usepackage{envmath}
\usepackage{natbib}
\usepackage{lineno}

\definecolor{orange-red}{rgb}{1.0, 0.27, 0.0}

\def\Sref#1{Sec.~\ref{#1}\xspace}
\def\Fref#1{Fig.~\ref{#1}\xspace}
\def\Tref#1{Table~\ref{#1}\xspace}
\def\Eref#1{Eq.~\ref{#1}\xspace}

\newcommand{\IUCAA}{Inter-University Centre for Astronomy and Astrophysics, Post Bag 4, Ganeshkhind, Pune 411 007, India}
\newcommand{\IPMU}{Kavli Institute for the Physics and Mathematics of the Universe (IPMU), 5-1-5 Kashiwanoha, Kashiwa-shi, Chiba 277-8583, Japan}
\newcommand{\WSU}{Department of Physics \& Astronomy, Washington State University, 1245 Webster, Pullman, WA 99164-2814, U.S.A.}

\begin{document}

\setlength{\parindent}{1em}
\setlength{\parskip}{.5em}




\title[]{SiGMa-Net: Deep learning network to distinguish binary black hole signals from short-duration noise transients} 

\author{Sunil Choudhary}
\affiliation{\IUCAA}

\author{Anupreeta More}
\affiliation{\IUCAA}
\affiliation{\IPMU}

\author{Sudhagar Suyamprakasam}
\affiliation{\IUCAA}

\author{Sukanta Bose}
\affiliation{\IUCAA}
\affiliation{\WSU}

\begin{abstract}

Blip glitches, a type of short-duration noise transient in the LIGO--Virgo data, are a nuisance for the binary black hole (BBH) searches. They affect the BBH search sensitivity significantly because their time-domain morphologies are very similar, and that creates difficulty in vetoing them. In this work, we construct a deep-learning neural network to efficiently distinguish BBH signals from blip glitches. 
We introduce sine-Gaussian projection (SGP) maps, which are projections of GW 
frequency-domain data
snippets on a basis of sine-Gaussians defined by the quality factor and central frequency. We feed the SGP maps to our deep-learning neural network, which classifies the BBH signals and blips. Whereas the BBH signals are simulated, the blips used are taken from real data throughout our analysis. We show that our network significantly improves the identification of the 
BBH signals in comparison to the results obtained using traditional-$\chi^2$ and sine-Gaussian $\chi^2$. For example, our network improves the sensitivity by 75\% at a false-positive rate of $10^{-2}$ for BBHs with total mass in the range $[80,140]~M_{\odot}$ and SNR in the range $[3,8]$. Also, it correctly identifies 95\% of the real GW events in GWTC-3. 
The computation time for classification is a few minutes for thousands of SGP maps on a single core. With further optimisation in the next version of our algorithm, we expect a further reduction in the computational cost. 
Our proposed method can potentially improve the veto process in the LIGO--Virgo GW data analysis and conceivably support identifying GW signals in low-latency pipelines. [This manuscript has been assigned the preprint number \textcolor{red}{LIGO-P2100485}.] 



\end{abstract}

\maketitle

\section{Introduction}


The age of Gravitational Wave (GW) astronomy has accelerated since the detection of the first GW event, GW150914~\cite{firstdetection,2016_transients}. As of now, the LIGO \cite{advligo} and Virgo \cite{Acernese_2015} collaborations have detected over 90 Compact Binary Coalescence~(CBC) signals \cite{gwtc1, gwtc2, gwtc2_1, gwtc3, 2020PhRvD.101h3030V, 2020ApJ...891..123N}, which include Binary Black Holes (BBHs), Neutron Star - Black Hole (NSBH) binaries and Binary Neutron Stars (BNSs).
Now, with KAGRA \cite{kagra}  joining the global network and with LIGO-India expected to come online later this decade~\cite{saleem2021science},
the rate of detection of
CBC events is expected to increase quite significantly.
One challenge that detection efforts face is erosion of some sensitivity of our detection pipelines caused by noise transients that share some characteristics with CBC waveforms~\cite{guide_ligo_noise}.
Some of these noise artifacts are 
difficult to differentiate from CBC signals, especially, 
when the latter are from binaries with 
large
total mass~\cite{Nitz_2018, 2018_dq_vetoes}. Considerable efforts have been invested in guarding against misclassification of those transients as CBC signals. This work contributes to that effort.

One specific type of short-duration noise transient, known as a blip glitch, is one of the major contributors to degrading the search sensitivity of CBC signals as well as other short-duration GW signals \cite{2016_transients, 2018_dq_vetoes}. Recent studies on blip glitches found that both 
LIGO detectors (located 
in Livingston and Hanford) report around 2 blip glitches per hour~\cite{Cabero_2019}. Blips are found in Virgo and GEO600 detectors as well~\cite{Cabero_2019}. The 
origin of 
a majority of blip glitches is still a mystery \cite{Cabero_2019}.  When it comes to the morphology of blip glitches, they are of short duration, $\mathcal{O}(10)$~ms in the time domain, and have a bandwidth of more than 100~Hz in the frequency domain.  
The shape of a blip glitch in the time domain resembles 
GW signals from 
BBHs 
either with
a large total mass, with highly asymmetric component masses, or with 
component
spins and orbital angular momentum anti-aligned. As a result, blips affect the efficiency of search of such BBH signals.  Studies conducted to 
diagnose
the source of blip glitches show that 
they 
have very little correlation with auxiliary channels. Only a few blips show correlations with the 
laser intensity stabilisation, computer errors ,power recycling control signals and relative humidity inside the detector.
In total, only around 8\% of LIGO Hanford (H1) 
and 2\% of LIGO Livingston (L1) blips in 
the
second
observation run (O2) run 
have shown correlations with any of the auxiliary channels~\cite{Cabero_2019}. 
It is possible that more than one physical mechanism is responsible for blips or that their origin may be quantum in nature. Hence, it is important to understand blip glitches and mitigate
their adverse effects on
the search sensitivity to
CBCs and other short-duration signals.
In this work, we have mainly used blip glitches identified by the Gravity Spy \cite{Zevin_2017, Gravityspy2, Gravityspy3, Gravityspy4} project from H1 and L1 LIGO O2 run and first part of third observational (O3a) run. Gravity Spy is a state-of-the-art tool to classify the glitches present in LIGO data. Up to now, it has identified 23 different types of glitches including blips in the LIGO data by combining efforts of human volunteers (citizen science) and deep learning networks.  

In the past, there have been efforts to understand blips~\citep[e.g.,][]{2013,PhysRevD.94.122004,datcharO3} and to mitigate their effect on CBC searches with a combination of modeling and statistical methods~\citep[e.g.,][]{Nitz_2018,PhysRevD.103.044035, 2020CQGra..37n5001D}. Even though some of these methods are effective at vetoing blips, there is still a lot of room for improvement. In recent years, machine learning (ML) algorithms have been proposed as being useful for GW data analysis. 
There have been several studies on classification, 
characterisation or parameter estimation of GW data with the help of ML algorithms such as Convolutional Neural Network (CNN), Recurrent Neural Network (RNN) and Variational AutoEncoders~\citep[e.g.,][]{PhysRevLett.120.141103,PhysRevD.104.064051,kapadia, magicbullet, rt_mma, nikhil_boosting, RNNsGW, aschmitt, 2020PhRvD.102f3015S,2018PhRvD..97d4039G,2021arXiv210403961Y,2018PhRvL.120n1103G}. CNNs are better suited for spatial data such as images~\citep[e.g.,][]{PhysRevD.104.064051, Zevin_2017} whereas RNNs work well with temporal or sequential data such as text or video \citep[e.g.,][]{RNNsGW, aschmitt}. Likewise, different ML algorithms can help in achieving various goals of GW data analysis. The purpose of this work is to exploit the tremendous image classification capabilities of CNNs in order to veto the blip glitches that appear in the LIGO--Virgo data.

This paper is arranged as follows. In \Sref{sec:SGmap}, we introduce the sine-Gaussian projection maps and explain how to generate them.  In \Sref{sec:datasim}, we give 
details of the simulations and data preparation. \Sref{sec:network} outlines the neural network that we constructed in this work. \Sref{existingchi} describes the existing methods like traditional $\chi^2$ and sine-Gaussian $\chi^2$ to veto blip like glitches. Finally, in \Sref{sec:results}, we present the results and performance of the network followed by \Sref{sec:conclusion} on conclusions and future prospects.

\section{Sine-Gaussian projection map}
\label{sec:SGmap}

\begin{figure*}

\includegraphics[width=1.0\textwidth]{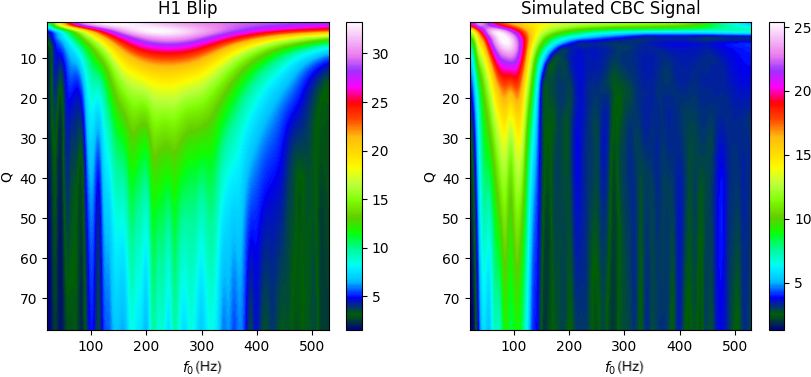}

\caption{Sine-Gaussian projection map of an H1 blip from O2 run with SNR 16 (left) and simulated BBH event+real noise with SNR 16 (right). Here
the
SNR for blips is calculated 
in the same way as
for BBH signals.
} 
\label{fig1}
\end{figure*}

A critical aspect of machine learning (ML) is presenting the data to the network in the most appropriate way. The data representation used in ML networks can have a significant impact on their performance.
There are machine learning studies where two-dimensional spectrogram of GW data like Omega Q-scan or Omicron Time-Frequency (TF) maps \cite{Robinet} have been used to categorize glitches~\citep[e.g.,][]{Zevin_2017} and  Continuous Wavelet Transform-based TF maps used to classify signals from glitches \citep[e.g.,][]{PhysRevD.104.064051}.
In this section, we introduce the sine-Gaussian projection (SGP) maps and explain how they are a 
useful way to represent GW data when it comes to distinguishing blips from BBH signals using a deep-learning image classifier.

Blips are one of the classes of glitches that resemble high mass-ratio or high component-mass BBH signals in the time domain and show similar TF Morphology \cite{Cabero_2019, Nitz_2018}. As a result, machine-learning algorithms that make use of TF maps may not be as efficient in distinguishing blips and high mass BBH signals. Alternatively, showing a contrast between the blips and high-mass BBH in the two-dimensional data by projecting these signals on sine-Gaussian parameter-space helps the machine learning network classify them well. Because the blips and high mass BBH signals show projection in different regions of the sine-Gaussian parameter-space represented by quality~factor~($Q$) and central frequency ($f_0$) and show easily distinguishable features, it is more suitable to use them for the classification problem. The projected
GW data snippet on the normalized sine-Gaussian waveforms 
in the $Q-f_0$ parameter-space is called an SGP map.
These SGP maps
constitute
our input data to the CNN. We call our neural network a \textbf{Si}ne \textbf{G}aussian projection \textbf{Ma}p-Convolution Neural \textbf{Net}work (\textbf{\textit{SiGMa-Net}}).


Mathematically, a sine-Gaussian waveform can be expressed as:
%
%
\begin{equation}
g(t)\equiv A e^{-4\pi f_0^2 \frac{(t-t_0)^2}{Q^2}}
\cos
(2\pi f_0 t +\phi_0) \,,
\label{sgexp}
\end{equation}
%
where $Q$ is the quality factor, $f_0$ is the central frequency, $\phi_0$ is the phase and $t_0$ is the central time of the sine-Gaussian waveform.
The projection of a data train $\mathbf{x}$ on the sine-Gaussian $\mathbf{g}$ is defined as \cite{PhysRevD.96.103018} 
\begin{equation}
\label{projection}
(\textbf{x},\textbf{g})=
4\Re \int_{f_{\rm{lower}}}^{f_{\rm{upper}}} \frac{\tilde{x}^{*}(f)\tilde{g}(f)}{S_{\rm{n}}(f)}df. 
\end{equation}
where
$S_{\rm{n}}(f)$ is the power spectral density of the noise,
and
$f_{\rm{lower}}$ and  $f_{\rm{upper}}$  are 
determined, respectively, by  the seismic cut-off frequency and the Nyquist frequency.

To construct an SGP map, we first choose the appropriate region of sine-Gaussian parameter-space onto which the data snippet is to be projected.
The choice of the parameter-space region depends on the kind of signal we are planning to project. 
We then sample points in the chosen region of the parameter space and data is projected over the sine-Gaussian waveform corresponding to each sampled point using \Eref{projection}. 
These calculated projections are represented using a color map,
where the color represents how strong or weak the
projection is,
as shown in \Fref{fig1}.
For our study, points are sampled uniformly in the region $f_0\in[20,520]$~Hz and $Q\in[2,80]$. The choice of these parameter ranges is made after extensive study of the projection of various blips and CBC signals. The pixels are smoothed using spline interpolation.

As we can see, there are 
clear
distinguishing features for blips and BBH signals in the SGP maps. Blip glitches show high projection at low $Q$ (in the range from 2 to 10) and have a projection for a wider range of frequencies than the BBH signals, as shown in \Fref{fig1}. The BBH signals, on the other hand, do not show much projection above 200~Hz. The extended projection along $Q$ depends on the component masses of the binary, although it is always more extended than blips along the $Q$ values. These projections of BBH signals are highly contrasting compared to the way noise projects on the SGP maps for high signal-to-noise ratio (SNR) signals (e.g., above match-filter SNR of 7). For low SNRs ($<7$), this contrast diminishes and identifying the BBH projections correctly becomes challenging. To overcome this issue 
impacting
low-SNR signals, we use 
multi-view learning, which is explained in \Sref{sec:datasim}.






\section{Data Simulation and Generation}
\label{sec:datasim}

\begin{figure*}[t]

\includegraphics[width=0.9\textwidth]{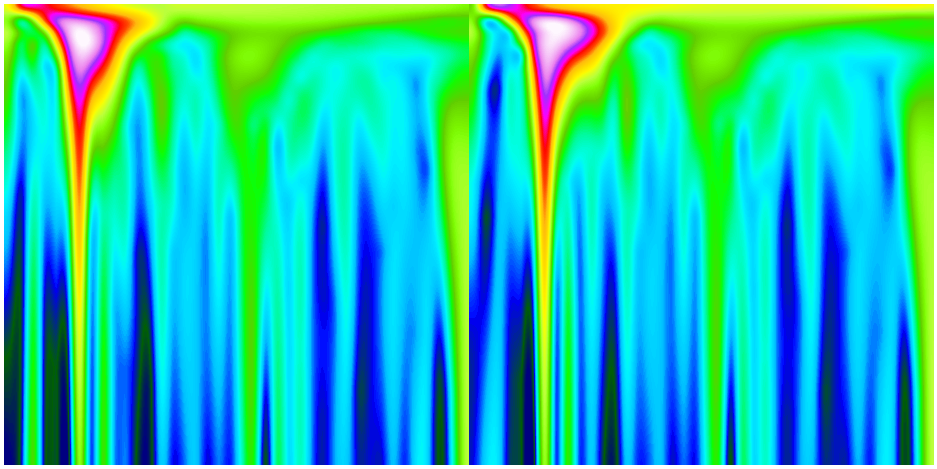}

\caption{Here we show an example of
a
simulated BBH input sample fed to the network. The H1 (left) and L1 (right) SGP maps are placed adjacent to each other. Axis labels and colorbars are omitted. Here, the SNR$\approx$10 for H1 and 
$\approx$9 for L1.   
} 
\label{net_bbh}
\end{figure*}

\begin{figure*}[t]

\includegraphics[width=0.9\textwidth]{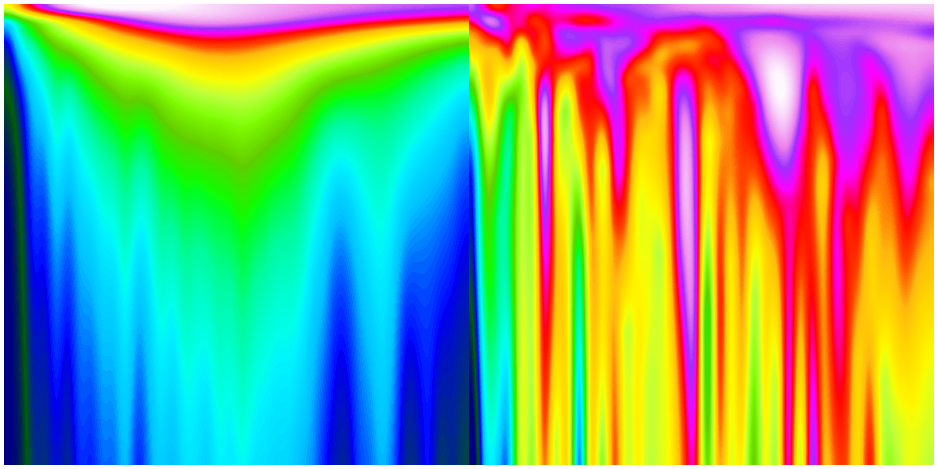}

\caption{Similar to \Fref{net_bbh}, 
except that a
{\em blip input sample} is fed to the network here. The H1 (left) and L1 (right) SGP maps are kept adjacent to each other. Axis labels and colorbars are omitted. We can see a blip is present in H1 (left) with an SNR$\approx$11. 
} 
\label{net_blip}
\end{figure*}

In this study, we work with real blips and 
simulated BBH signals injected in real data.
We describe below the procedure followed for fetching the real data and generating the simulated data in the form of a time-series which are then used to prepare the SGP maps.

We identify GPS times 
of
the blip glitches from
Gravity Spy and then fetch the corresponding 
H1 and L1
data from O2 and O3a 
\cite{2021_gwosc}. While most of our blips are
taken from O2, we choose 30\% of the 6000 test sample blips from O3a. 
We use the ``cleaned C02" and ``cleaned C01" frame types for the O2 and O3 data, respectively~\cite{datcharO3}. For each blip, we choose a 16~sec-long data segment out of which a 2~sec-long segment is chopped such that the GPS time falls at the center of the segment while the whole data segment is used to calculate the power spectral density (PSD), as described later in this section. 

We simulate non-spinning BBH signals using the \textsc{IMRPhenomPv2} \cite{PhenomP} template such that their masses $m1, m2\in[10,100]~M_{\odot}$ and their sky locations follow a uniform distribution. The frequency cutoff on the lower end  ($f_{\rm{lower}}$) is set to 20~Hz and the upper cutoff ($f_{\rm{upper}}$) is set to 2048~Hz. The match-filtering SNR for BBHs is also drawn uniformly from the range [4,10] for the H1 detector. This fixes the SNR for the BBH signal in the L1 detector, accordingly. We inject the simulated BBH signals into the real O2 noise as explained below. First, we identify a 64~sec-long data segment from the O2 data which has no known triggers. We further verify by plotting a Q-Transform map and ensure that this data segment has no GW signals and is artifact-free. Next, we randomly draw 2~sec-long data segments from this main segment such that the peak of the BBH signal is located at the center of the segment.

Subsequently, we apply a windowing function to the BBH and blip time-series data samples in order to reduce the spectral leakage due to discontinuity at the edges of the chopped segment. Windowing ensures a smooth and gradual transition of time-series amplitude to zero at the edges. 
As the input to our network is the SGP maps, it is crucial to find an optimal combination of various factors that will maximise the projection of BBH signals and blips. As a result, the length of the data segment, the location of the signal or blip within the segment and the choice of windowing had to be tuned by performing several tests. For example, longer data segments result in increased noise in the SGP maps and shorter data segments also adversely affect the SGP maps due to the windowing process. Hence, we find that an intermediate length of two seconds is the most appropriate choice for the signal parameter ranges chosen for this work.

Once the time series data for blips and BBH are prepared, we proceed to generate the SGP maps for our sample. The first step is to calculate the PSDs. 
For each blip, we use the 16~sec-long data segment which contains the blip and apply the standard Welch's method to calculate the PSD. 
This method ensures that the presence of the blip in the segment does not affect the PSD calculation. For the BBH sample, we use the same 64~sec-long data segment which was used as the background noise for the BBH signals. We apply the same Welch's method to get the PSD required for the BBH sample. 


In the next step, we calculate the projection of the data (BBH and blips) on the sine-Gaussian waveform (see \Eref{sgexp} and \Eref{projection}) and represent it using the 2D parameter-space of $Q-f_0$ (see \Sref{sec:SGmap} for details) for each detector separately. The color in each pixel of an SGP map represents the strength of the projection.

Initially, we experimented with analysing SGP maps of individual detectors separately for each blip and BBH signal. While the network performance was satisfactory at high-SNRs, there was room for further improvement in the sensitivity of low-SNR GW events. We then considered analysing maps from both detectors simultaneously. There is a clear advantage of working with multi-detector data because a true astrophysical signal such as that from a BBH, if sufficiently strong, will appear in both the detectors whereas non-astrophysical noise transients such as blips rarely coincide for two independent detectors. This is well evident for an example data of BBH (\Fref{net_bbh}) and a blip (\Fref{net_blip}) showing both H1 and L1 detectors. One could consider including data from more than two detectors which may or may not be efficient in our end goal. However, we leave this exercise for future work and stick to using two-detector data.

In order to analyze SGP maps from both detectors simultaneously, 
we consider a multi-view learning model of the CNN e.g., \cite{2017arXiv170500034B}. In this form of learning, the network learns using multiple image representations of a given class of objects. The multi-view model of learning increases the distinguishable features in the input images and the learning capability of the network as a whole.
Multi-view learning could be done in mainly two ways, parallel-view and merged-view. In parallel-view learning, images are fed to the network through more than one input channel and combined subsequently after passing through a few hidden layers. On the contrary, in merged-view learning, the images are placed adjacent to each other in a grid and then fed to the network through a single input channel. 

In this study, we adopt the merged-view learning where the SGP maps, for each BBH or blip, created using the H1 and L1 detectors, are placed laterally as a single input image for the network. For example, see \Fref{net_bbh} and \Fref{net_blip} where the images, without any labels and tick marks, are fed to the network.  

We choose 10000 SGP maps for training and 3000 SGP maps for validation where the blips and BBH signals are in equal proportion. 
For the test sample containing BBH signals, we produce two distinct samples as a function of mass and mass-ratios. Each BBH sample comprises 12000 SGP maps. We produce 6000 SGP maps of blips for the test sample. 
Additionally, we use the sample of 49 real GW events taken from O1, O2 and O3a~\cite{gwtc1,gwtc2}.

\section{Deep learning network}
\label{sec:network}
 

 
 
 
 
 
 
 
 
 
The goal of our study is to distinguish BBH from blip glitches.
For this classification problem, we use a deep learning algorithm known as Convolutional Neural Network \cite{cnnidea,li2020survey,DL_overview}. We follow the supervised learning approach where we train the network on SGP maps belonging to both the BBH and blip classes. The trained network then gives us predictions whether a test SGP map contains a BBH or a blip. 

\begin{figure*}[t]
\includegraphics[width=1.0\textwidth]{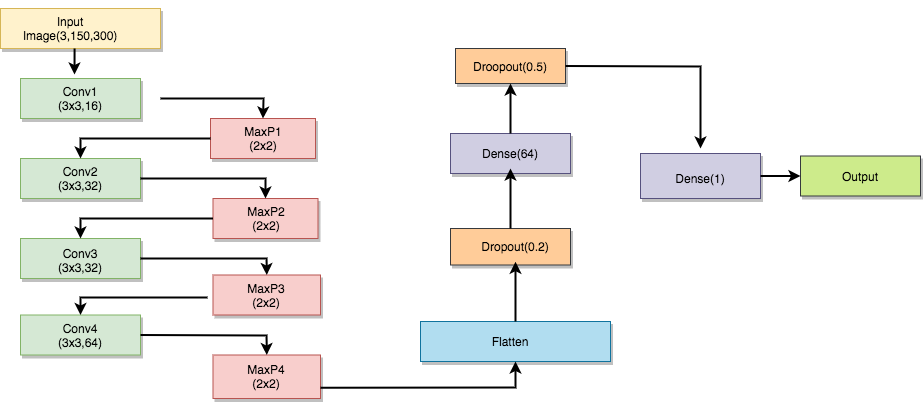}

\caption{Schematic diagram of our neural network. Each layer uses \emph{relu} (rectified linear unit) activation function except for the last one, which uses a sigmoid function. 
The final output is the probability that the input image corresponds to a BBH signal.
Here, Conv$k$ stands for $k$th convolutional layer, MaxP$k$ stands for $k$th max-pooling layer, Flatten denotes the flattening of 2D matrix, Dropout($x$) layer is to tackle overfitting by dropping some of the neurons and Dense($k$) is fully connected layer with $k$ neurons~\cite{upgrad}.} 
\label{CNNfig}
\end{figure*}

We describe below the structure of the network developed in this work and summarise the training process leading to the final model that will make robust predictions. As mentioned in the previous section, the input data for the network are two SGP maps corresponding to H1 and L1 detectors for each BBH or blip. These maps are placed laterally such that their final dimension is 150$\times$300 along with 3 color channels as seen by the network.
Our network has four convolution blocks followed by a flattened layer and two fully connected blocks which are connected to an output layer (see \Fref{CNNfig}). Each convolution block has a convolution layer and a pooling layer. We choose the \emph{relu} (rectified linear unit) activation function in the convolution layers and a max-pooling option in the pooling layer. Max-pooling has been shown to perform better in classification tasks where one has to deal with sharp features in the images in comparison to other options such as average-pooling and min-pooling.
Subsequently, we introduce a flattened layer followed by the relu activation function. Next, in each of the fully connected blocks, we have a Dropout and a Dense layer. The dropout layer helps in preventing over-fitting issues. Finally, in the output layer, we apply the sigmoid activation function to predict the classification probabilities. This layer has a single neuron which gives the probability that the input image belongs to BBH class. In order to determine the probability for the class of blips, we simply subtract the predicted probability from 1. 

We choose binary cross-entropy to define the loss-function of the network as it is appropriate for binary classification. We use the ``\emph{adam}" optimizer \cite{kingma2017adam} with a learning rate set to 0.001 and other parameters set to their default values as they happened to be an adequate choice based on the tests we conducted. We use the metric ``accuracy" to measure the network performance. 

Our training sample includes 5000 maps of blips and BBHs each whereas the validation sample includes 1500 maps of blips and BBHs each. We train our network for 10 epochs and each epoch consists of the training samples divided into 300 batches. The validation sample is divided into 100 batches. During the training process, we use the ``fit generator" method available in \textsc{Keras} \cite{2015keras} which loads the data into primary memory in batches and feeds it to the network for training. As we do not load the whole training data at once into the memory, it reduces the load on the computing node and makes the training process quite fast.

\section{Existing methods to tackle blips}
\label{existingchi}

Before testing the performance of our machine learning model and comparing it with existing methods for vetoing blips in GW data, we briefly summarize two of the main statistics in the GW data analysis which are currently employed for vetoing the blips, namely, traditional $\chi^2$ \cite{b_allen} and sine-Gaussian $\chi^2$ \cite{Nitz_2018}.

The traditional $\chi^2$ is constructed by sub-dividing the 
template waveform triggered by the match-filtering process into $p$ non-overlapping frequency bins. Each bin contributes equally to the SNR of the best matching CBC template~\cite{b_allen}. If the data segment $\mathbf{s}$ is adequately described as a Gaussian noise with an added CBC signal that shows a large correlation with the template $\mathbf{h}$, this will follow a reduced $\chi^2$ distribution with $2p-2$ degrees of freedom. Most noise transients present in the GW data show a higher $\chi^2$ in comparison to CBC signals, making it easy to distinguish signals vs noise artifacts.  
The definition for the traditional $\chi^2$ is   
\begin{equation}
    \chi^2_{\rm{r}}=\frac{1}{2p-2} \sum_{i=1}^{p} ||\langle \textbf{s}|\textbf{h}_i\rangle-\langle \textbf{h}_i|\textbf{h}_i\rangle||^2 ,
\end{equation}
%
where $p$ is the number of bins, $\textbf{h}_i$ is the time domain representation of the waveform corresponding to
the
\emph{i}th bin, and for any two data segments $\mathbf{a}$ and $\mathbf{b}$ the inner product is 
\begin{equation}
    \langle \textbf{a}|\textbf{b}\rangle=4\int_0^{\infty} \frac{\tilde{a}(f)\tilde{b}^*(f)}{S_{\rm{n}}(f)} df .
\end{equation}
The traditional $\chi^2$ is then combined with the SNR to create a ranking
statistic
called re-weighted SNR, 
which
as used in the \textsc{PyCBC} \cite{pycbc_soft, 2016_pycbc_usman, 2020PhRvD.102b2004D,PhysRevD.85.122006} analysis, is:
\begin{equation}
    \tilde{\rho}=
     \begin{cases}
     \rho & \text{for $\chi^2_{\rm{r}}\leq1$}\\
     \rho \Big[ \frac{1}{2}(1+(\chi^2_{\rm{r}})^3)\Big]^{-\frac{1}{2}} & \text{for $\chi^2_{\rm{r}}>1$}
     
     \end{cases}
     .
\end{equation}

The sine-Gaussian (SG) $\chi^2$  takes 
advantage of the excess power that blips typically possess in higher frequency regions in comparison to CBC signals. It computes this excess power by utilizing a set of sine-Gaussian waveforms whose central frequency ranges from 30-120~Hz above the final frequency of the triggered template~\cite{Nitz_2018}. It is defined as
\begin{equation}
    \chi^2_{\rm{r,sg}}=\frac{1}{2N}\sum_{i=1}^N 
    \langle \textbf{s}|\textbf{g}_i \rangle^2,
\end{equation}
where $\textbf{g}_i$ is \emph{i}th sine-Gaussian waveform.

The re-weighted SNR used in the case of SG $\chi^2$ is given as 
\begin{equation}
    \tilde{\rho_{\rm{sg}}}=
     \begin{cases}
     \tilde{\rho} & \text{for $\chi^2_{\rm{r,sg}}\leq4$}\\
     \tilde{\rho}(\chi^2_{\rm{r,sg}}/4)^{-\frac{1}{2}} & \text{for $\chi^2_{\rm{r,sg}}>4$ }
     
     \end{cases}
     .
\end{equation}
The number of bins $p$ in the case of the traditional $\chi^2$ and the number of sine-Gaussian waveforms in the case of the SG $\chi^2$ 
are tuned empirically based on the distribution of single detector background triggers in engineering data.

\section{Performance and Results}
\label{sec:results}

\begin{figure*}[t]

\includegraphics[width=1.0 \textwidth]{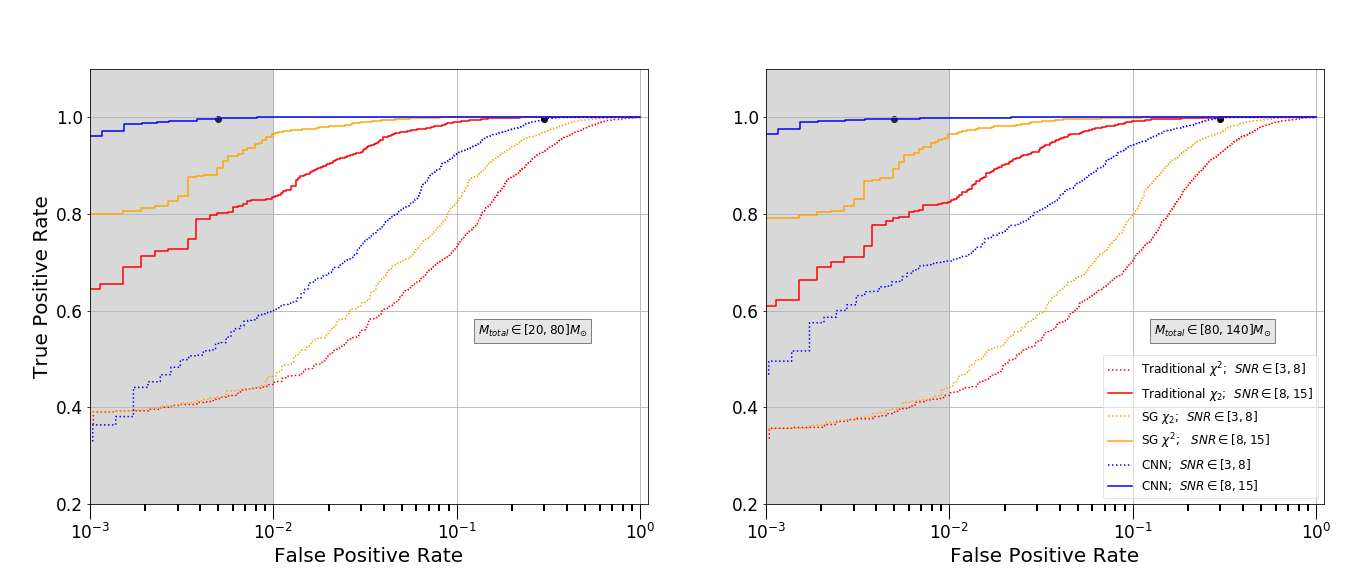}

\caption{Comparison of ROC curves for CNN 
with 
those for two
other methods. The curves are shown for low- and high-SNR bins (solid and dotted, respectively). On the left, we show the performance at low-mass BBH signals, which is similar to
that of
the high-mass BBH signals on the right. The grey shaded region has more uncertainty for the low SNRs (dotted curve) due to limited sample size. The solid circle marks the TPR-FPR for a threshold of 0.8.}   
\label{fig:ROC1}
\end{figure*}

\begin{figure*}[t]
\includegraphics[width=1.0 \textwidth]{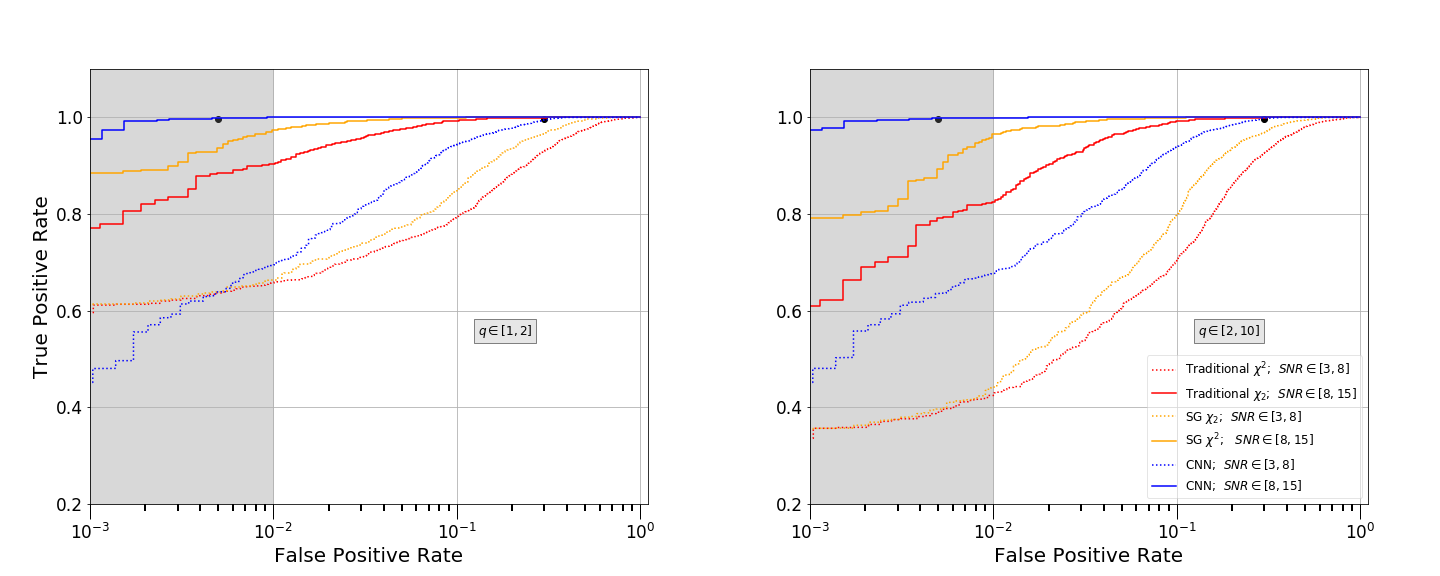}
\caption{ Comparison of ROC curves for CNN with
those for two
other methods. Same as \Fref{fig:ROC1} except the performance is shown for low and high mass-ratios of BBH signals (left and right, respectively).}
\label{fig:ROC2}
\end{figure*}

\begin{table}[t]
\centering
 \begin{tabular}{p{3.5cm} c c p{1.5cm} } 
\hline
Type  &  \multicolumn{2}{c}{SNR}  & Total       \\
      &   [3,8]  &  [8,15] & \\
\hline
\hline
Blips  &  3000 & 3000  & 6000 \\   
\hline
BBH1  & & & 12000\\ 
 $M_{\rm{total}}\in[20,80]~M_{\odot}$     &  3000 & 3000  & \\
 $M_{\rm{total}}\in[80,140]~M_{\odot}$      &  3000 & 3000 & \\
\hline
BBH2 & & & 12000\\   
 $q\in[1,2]$  &  3000 & 3000  &\\
 $q\in[2,10]$  &  3000 & 3000  &\\
\hline
GWTC-2 and GWTC-3 &  & & 81 \\ 
\hline
Other marginal and 4-OGC events & & & 34\\
\hline

\end{tabular}
 \caption{ Details of the Test Sample. Whereas blips are 
  divided
 based on their SNRs alone, the BBH samples are divided based on their SNRs and 
 a mass parameter, which is taken to be
 either
 their total mass or their mass-ratio. The total number of unique samples for each type is given in the last column. }
\label{tab:testsamp}
 \end{table}

After training our network on a sample of real blips and simulated BBH signals, we obtain a final model which can make predictions on a test sample.  
In this section, we evaluate our model's performance using some of the standard metrics, for example, the receiver-operating characteristic (ROC) curves and F1-score. We also present model predictions for the real GW events.

First, we describe how the test sample is prepared.
We produce a sample of 6000 blips which are divided equally into two SNR bins. 
For the BBHs, we generate two distinct samples such that one has a uniform distribution in total mass and the other is uniform in mass-ratio. Each of the BBH samples is divided equally into two bins as per the total mass or the mass-ratio. These sub-samples are further divided into two SNR bins. The breakdown of the full test sample according to the bins, how the bins are defined and the unique number of test samples used in our analysis are given in \Tref{tab:testsamp}. We also list the number of GW events tested in this work.



Next, we test the performance of our network in each bin for BBH versus blips. To generate the ROC curves, we calculate the true-positive rate (TPR) and false-positive rate (FPR) for varying thresholds of detection. The resulting ROC curves for BBH1 sample split by their total mass are shown in \Fref{fig:ROC1} and for BBH2 sample split by their mass-ratio are shown in \Fref{fig:ROC2}. As expected, the network performs better at high SNRs (solid curves) than low SNRs for both BBH1 and BBH2 samples. The grey band roughly shows the region where we have limited samples in low SNR bins which increase the uncertainties on the ROC. Hence, for the low SNRs, we focus on FPR$>0.01$. Interestingly, the ROC curves are only marginally different across the bins in total mass or the mass-ratio. It is encouraging to see that the network is robust and performs equally well in the total mass or mass-ratio ranges we explore here.

\begin{figure}
\includegraphics[width=0.5 \textwidth]{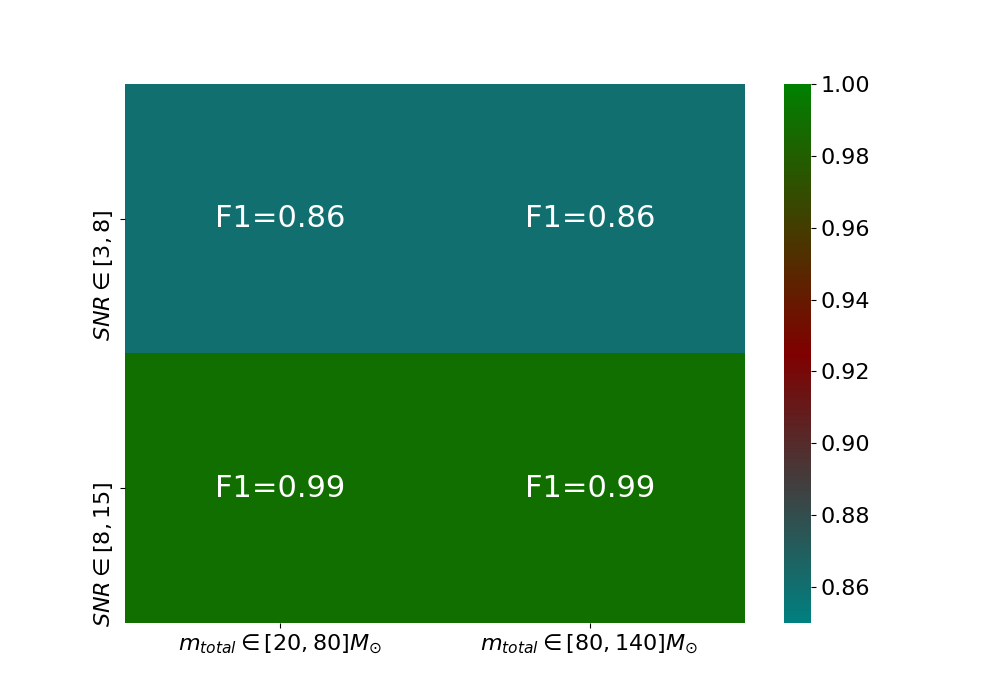}
\caption{F1 score of our neural network for a threshold of 0.8. Scores are shown for low and high total-mass of the BBH signals (x-axis) further split by their SNRs (y-axis).}
\label{fig:F1}
\end{figure}

\begin{figure}
\includegraphics[width=0.5 \textwidth]{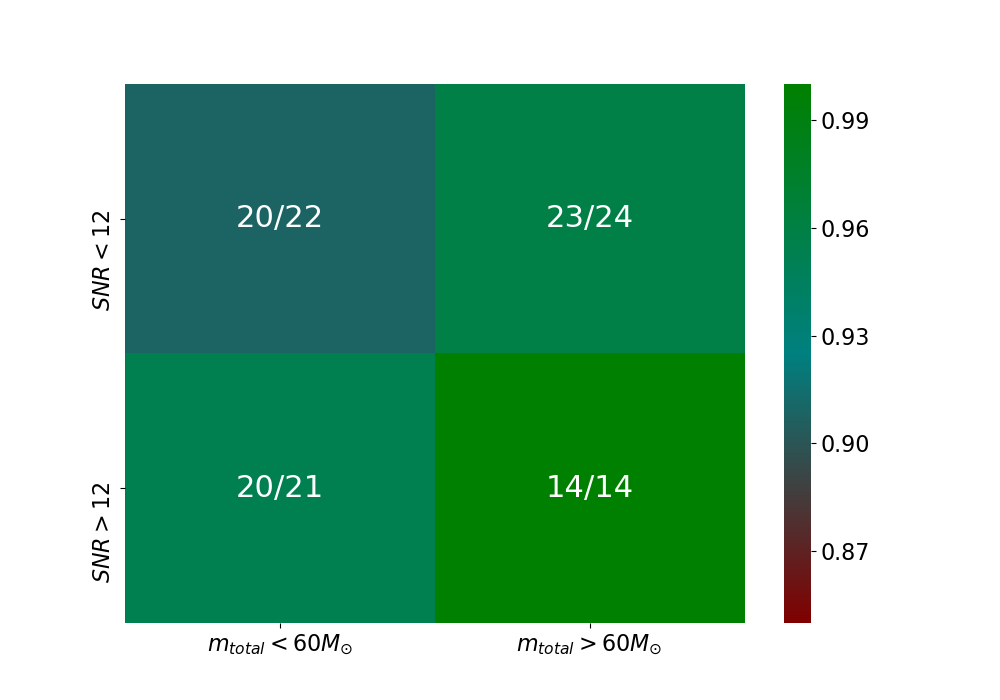}
\caption{Detection efficiency of our network on the real BBH events from GWTC-3. The format is the same as in \Fref{fig:F1} and the threshold is also set to 0.8.}
\label{fig:realBBH}
\end{figure}

We also compare the performance of the neural network classifier with existing methods to veto blips mentioned in \Sref{existingchi}, namely traditional $\chi^2$ and SG $\chi^2$.
We can see from the ROC curves that our network performs better than the traditional $\chi^2$ and SG $\chi^2$, which are used here in their network form calculated as quadrature sum of their value in H1 and L1 detectors. Although, in current LIGO search pipelines these statistics are applied in modified form with improved modeling of background distribution \cite{improved_pycbc, 2018PhRvD..98b4050N, 2020PhRvD.102b2004D}. The performance of our neural network classifier is better at higher total masses and higher mass ratios  in comparison to the lower bins. 
In all cases, the CNN performs either better or comparable to traditional $\chi^2$ and SG $\chi^2$ . 
It is important to highlight the improvement in sensitivity with our network at low SNRs for high total mass and/or high mass-ratio BBHs where the existing methods are known to have a poor performance. 
At high masses, our network shows 75\% increment in the TPR than the traditional $\chi^2$ and SG $\chi^2$ at an FPR of $10^{-2}$ and for high mass-ratios, the neural network shows around 50\% improvement in TPR at an FPR of $10^{-2}$ compared to other methods.    

We use another metric called F1-score to assess the efficiency of our network.
The F1-score is calculated for the two SNR and two total mass bins. The F1 score values are shown in \Fref{fig:F1}. As we can see, the network performs well both at low and high SNRs. The performance seems to be almost independent of the total mass of the binary. 


We also test our network on the real events from GWTC-3 and the results are shown in \Fref{fig:realBBH}. Four events are not included in our sample, as one of the detector(H1 or L1) data is not available for these events. 
The network predicts 95\% GW events correctly as BBH signals. Among the high SNRs, the event GW170817 is misclassified, which is a binary neutron star merger signal with a glitch overlapping in the L1 detector. 
Among the low SNR events, GW191219\_163120, GW200308\_173609 and GW200220\_124850 are not classified as BBH. Upon inspection of the SGP maps, we find that there is a hint of projection in one of the detectors but the projection in the other detector is not very clean due to low SNR. This is consistently seen in the maps of all of the three events which is the most likely explanation for the mis-classification.   


In \Tref{tab:comptimes}, we show the computation times for analysing GW data with our network.  
The training and classification of several thousand samples by the network is quite fast even on a single core. Although the generation of SGP maps is done using multiple cores, which makes it reasonably fast, there is further room for improvement in the process of generating a single SGP map.
The latter process executes on a single core currently. We anticipate that through some changes, for example, parallellization of certain loops and by reducing the data sampling rate from 4096 to 2048~Hz without affecting our results, we can speed up the generation of an SGP map by factor of a few.

\begin{table}[t]
\centering

 \begin{tabular}{p{4.5cm} | p{3.cm}} 
\hline
Action & Time \\
\hline
\hline
Generating 1 SGP map & 30 sec  \\ 
\hline
Training of the network on 10000 samples &  20 min\\ 
\hline
Classification of 3000 SGP maps & 90 sec \\ 
\hline
\hline
\end{tabular}
 \caption{ Computation time for various sections of SiGMa-Net. All times are computed for single core of \emph{Intel Xeon  Gold 6142} CPU. Multiprocessing and parallelizing further reduces the computation times. For example, parallelizing on N processors reduces all computational costs by a factor of N.   }
\label{tab:comptimes}
 \end{table}

\section{Conclusion and Future Prospects}
\label{sec:conclusion}

Blip glitches
are known to have similar time-frequency characteristics to those of 
BBH signals of high total mass. Thus, the standard pipelines find it difficult to distinguish high-mass BBH signals from blips, in particular.  
We use sine-Gaussian projection (SGP) maps, 
a new way of visualising  GW data, to discriminate BBH signals from blips. We create the SGP maps by projecting the BBH signals and blips on the two-dimensional parameter space of sine-Gaussian waveforms. These maps are complementary to the continuous-wave transform maps commonly used in GW data analyses.

In order to demonstrate the usefulness of SGP maps we develop a deep learning framework that uses convolutional neural network to classify BBH signals and blips. It can help in reducing the ambiguity between CBC and Blip triggers generated from LIGO--Virgo runs. We compare our method with those used in the standard pipelines for identifying BBH signals from the LIGO--Virgo GW data.
We find that our network performs consistently much better than traditional $\chi^2$ and sine-Gaussian $\chi^2$ for BBH signals with SNR $>8$ and a total mass $\in[20,140]~M_{\odot}$ where the mass ratio goes from 1 to 10.
For low SNRs ($\in[3,8]$) too, the network performs significantly better when FPR$>0.01$.  However, we notice that below FPR of $\approx$~0.01, the limited size of our training sample does not allow characterisation of our network robustly as we do not have sufficient numbers of blips in real data from first two observing runs. This will not be a limiting factor in our subsequent work where we plan to include data from third observing run as well.   
We also find that our network is able to correctly identify 77 out of 83 CBC events from LIGO--Virgo's O1, O2 and O3 observing runs.  Our network correctly identifies 15 out of the 19 LIGO marginal events \cite{gwtc3} and all of the 15 GW events from the 4th Open Gravity Catalogue \cite{nitz20214ogc}.

Although, this work specifically focuses on blips which is a sub-category of glitches found in LIGO--Virgo data, in the future, we plan to include other types of glitches and construct a more generic neural network to  detect CBC signal with the help of SGPs. 
We could also consider including data from additional detectors (e.g., Virgo and KAGRA). Inclusion of multiple SGP maps for the same BBH will most likely improve the sensitivity of identifying them but at the cost of increased computational times as the network will have to process data. However, more detector data also increases the probability of finding coinciding glitches. This could then adversely affect the performance of the neural network. Thus, one will have to do a risk-benefit exercise before deciding whether or not to use data from additional detectors which we leave to a future study. 

In the future, with the inclusion of various artefact categories in the training, SiGMa-Net could be developed to potentially analyze raw GW data in low-latency searches to generate an initial set of triggers as candidate BBH events where the speed would be significantly higher than the match-filtering based template bank search.

\section{Acknowledgment}
We are thankful to Khun Sang Phukon and Thomas Dent for their helpful comments and suggestions. We would also like to thank our colleagues Shreejit Jadhav and Sanjit Mitra at Inter-University Centre for Astronomy and Astrophysics, Pune, India (IUCAA) for useful discussion. This material is based upon work supported by NSF's LIGO Laboratory which is a major facility fully funded by the National Science Foundation. The computation platform for this work was provided by Sarathi cluster at IUCAA, which is part of the LIGO scientific collaboration data grid facility. We would also like to acknowledge help and support of their staff specially Deepak Bankar for timely help related to Sarathi cluster issues.
SS is supported by Navajbai Ratan Tata Trust (NRTT) and LIGO - India grants at IUCAA.

\bibliography{ML.bib}

\end{document}